\documentstyle[aps,prl,twocolumn]{revtex}
\input epsf

\begin{document}
\draft
\title{The Experimental Observation of a Superfluid Gyroscope in a dilute Bose Condensed Gas.}
\author{E. Hodby, S.A. Hopkins, G. Hechenblaikner, N.L. Smith and C.J. Foot}
\address{Clarendon Laboratory, Department of Physics, University of Oxford,\\
Parks Road, Oxford, OX1 3PU, \\
United Kingdom.}
\date{\today}

\maketitle

\begin{abstract}
We have observed a superfluid gyroscope effect in a dilute gas
Bose-Einstein condensate. A condensate with a vortex possesses a
single quantum of angular momentum and this causes the plane of
oscillation of the scissors mode to precess around the vortex
line. We have measured the precession rate of the scissors
oscillation. From this we deduced the angular momentum associated
with the vortex line and found a value close to $\hbar$ per
particle, as predicted for a superfluid.

\end{abstract}

The superfluid nature of a dilute gas Bose condensate, most
strikingly demonstrated by its response to rotation, is currently
an area of great experimental interest. The unique response of a
superfluid to an applied torque arises because the presence of its
single macroscopic wavefunction constrains the flow patterns
allowed within its bulk. In \cite{Madison}, and later in
\cite{Ketterlee,Hodby}, vortices with quantised circulation were
observed when the trapping potential confining the atoms was
rotated above a critical speed. Such vortices correspond to phase
singularities within the gas where the density goes to zero. The
quantised circulation (and hence angular momentum) associated with
each vortex ensures that the superfluid wavefunction is
single-valued at all points, whilst enabling the condensate to
mimic the rotating flow pattern of a normal fluid. Secondly, the
observation of the oscillation frequency of the 'scissors mode'
\cite{Marago} gave further evidence for superfluidity by showing
that only an irrotational flow pattern (and not a rotational one)
was possible in a vortex-free condensate.

\begin{figure}
\begin{center}\mbox{ \epsfxsize 2in\epsfbox{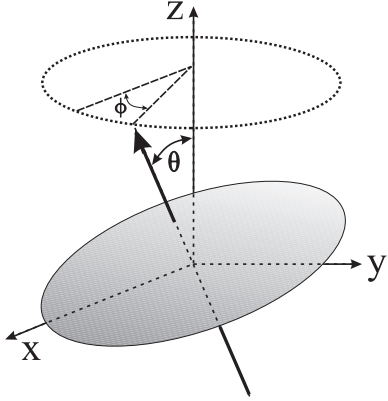}}\end{center}
\caption{The gyroscope motion. The condensate performs a fast
scissors oscillation in the $\theta$ direction at $\omega_{sc}$,
whilst the plane of this oscillation slowly precesses in the
$\phi$ direction at frequency $\Omega$. The vortex core moves
independently and is not shown on this diagram. }\label{gyro}
\end{figure}

In this work, we describe the realisation of a superfluid
gyroscope as discussed by Stringari in \cite{Stringari}, in an
experiment that combines both vortex production and the excitation
of the scissors mode. The scissors mode is a small-angle
oscillation of the condensate relative to the trap potential, that
is excited by a sudden tilt of the trap, as described in detail
elsewhere \cite{Marago,GOS}. In the gyroscope experiment we excite
the scissors mode oscillation in the XZ or YZ plane of an axially
symmetric condensate containing a vortex along the Z axis. In the
presence of the vortex the plane of oscillation of the scissors
mode precesses slowly around the Z axis. In polar coordinates, the
scissors oscillation is in the $\theta$ direction and the
precession is in the $\phi$ direction as shown in fig.~\ref{gyro}.
From the precession rate we deduce the angular momentum associated
with the vortex line $\langle L_z \rangle$ and hence show that
this angular momentum is quantised into units of $\hbar$ per
particle, as predicted for a superfluid.

The relationship between the precession rate $\Omega$ and $\langle
L_z \rangle$ may be derived by considering the scissors mode as an
equal superposition of two counter-rotating $m = \pm \,1$ modes
\cite{string}. These modes represent a condensate tilted by a
small angle from the horizontal plane rotating around the z axis
at the frequency of the scissors oscillation, $\omega_{\pm}$ =
$\omega_{sc}$. The symmetry and hence degeneracy of these modes is
broken by the presence of axial angular momentum $\langle L_z
\rangle $. Provided that the splitting is small compared to
$\omega_{sc}$ we observe a precession of the scissors mode in good
agreement with the theory of \cite{Stringari}:
\begin{equation}
\Omega = \frac{\omega_+ - \omega_-}{2} = \frac{ \langle L_z
\rangle }{2 m N \langle x^2 + z^2 \rangle }
 \label{first}
\end{equation}
where $N$ is the total number of atoms in the condensate and $m$
is the atomic mass. Substituting for $ \langle x^2 + z^2 \rangle $
in the case of a harmonically trapped condensate one obtains
\cite{Stringari,Svid}

\begin{equation}
\Omega = \frac{7 \omega_{sc}}{2} \frac{\langle L_z \rangle}{N
\hbar} \frac{\lambda^{5/3}}{(1+\lambda^2)^{3/2}} \left(15 N
\frac{a}{a_{ho}} \right)^{-2/5}
 \label{precession}
\end{equation}
where $\lambda = \omega_z / \omega_{\perp}$, $a_{ho} = (\hbar / m
(\omega_{\perp}^2 \omega_z)^{1/3})^{1/2}$ and $\omega_{sc} =
(\omega_x^2 + \omega_z^2)^{1/2}$.

Other superfluid gyroscopic effects have been observed but the 3 -
dimensional interaction between the velocity field of the vortex
and the scissors mode of the condensate make this experiment
unique. Superfluid gyroscopes of liquid helium exhibit persistent
currents in toroidal geometries, with many quanta of circulation
\cite{He}. In contrast, we show here that a single vortex of
angular momentum $ \langle L_z \rangle = N \hbar$ significantly
modifies the motion of a trapped BEC gas in an excited state. This
is possible because the vortex produces relative shifts in the
excitation spectrum of order $\xi / R_0$ and in such a dilute
system the vortex core size $\xi$ cannot be ignored with respect
to the average size of the condensate $R_0$ \cite{Svid}. Related
experiments with vortex lines in dilute trapped gases are
described in \cite{Chevy,Haljan}. The angular momentum of a vortex
line was measured in \cite{Chevy} using the precession of a radial
breathing mode (a superposition of $m=\pm \, 2$ quadrupole modes)
in the plane perpendicular to the vortex line. In that work,
motion is confined to 2 dimensions and the quadrupole oscillation
of the condensate does not affect the vortex line. We observe the
precession of a different quadrupole mode and the motion of the
vortex line relative to the condensate is of interest. It has been
suggested in [8] that the vortex line might follow the bulk motion
of the condensate, and we present evidence consistent with that
view. In \cite{Haljan} the precession of a vortex line is observed
in the absence of any bulk condensate motion, when it is tilted or
displaced from the condensate symmetry axes.

It is worth noting that a 'gyroscope' is a general term,
describing a frictionless system with a large angular momentum
vector that is able to rotate about any axis \cite{OED}. Our
condensate, supported in a frictionless magnetic trap and with a
vortex, is an example of such a system, although one must be
careful not to draw incorrect analogies to classical gyroscope
systems. Nutation, the wobbling motion superimposed on the
precession of a spinning top is not analogous to the scissors
motion in our system. The nutation frequency depends on the
angular momentum, whereas the scissors frequency does not
\cite{Kleppner}.

The first stage of exciting the superfluid gyroscope is to
nucleate a single vortex at the centre of a condensate. A detailed
discussion of the conditions for vortex nucleation in our
apparatus is given in \cite{Hodby}. In summary the excitation
procedure used for this experiment was as follows: First we
produced a condensate in an axially symmetric TOP trap with
$\omega_{\perp} / 2 \pi = 62\,$Hz and $\omega_z / 2 \pi = 175
\,$Hz. To spin up the condensate we made the trap eccentric
$(\omega_x / \omega_y = 1.04)$ over 0.2 seconds, with a trap
rotation rate of $44 \,$Hz. After holding the condensate in the
spinning trap for a further 1 s we ceased the rotation of the trap
potential by ramping both the trap rotation rate and the trap
eccentricity to zero over 0.4 s. During the whole vortex
excitation process, the RF evaporation was kept on, maintaining
the temperature at approximately $0.5 \, T_c$.

\begin{figure}
\begin{center}\mbox{ \epsfxsize 2in\epsfbox{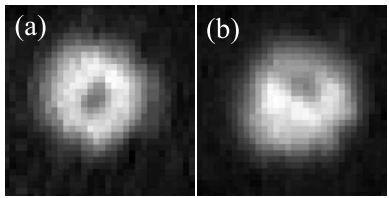}}\end{center}
\caption{Expanded images of vortices prior to excitation of the
scissors mode, viewed along the axis of rotation. (a) shows a
single centred vortex whilst (b) shows a vortex at the edge of our
criterion for an `acceptably centred vortex'.}\label{toppic}
\end{figure}

Figure~\ref{toppic} shows a perfectly centred vortex (a) and one
that is at the edge of our criterion for an acceptably centred
vortex (b). The position of the vortex line is important because
we use destructive imaging; each data point requires a new
condensate and identical starting conditions to give the same
precession rate for each run. Equation~\ref{precession} shows that
the precession rate is affected by the number of atoms in the
condensate and the angular momentum associated with a vortex line.
Our shot-to-shot number variation of $N = 19000 \pm 4000$ produces
a $10 \%$ variation in the precession rate. More significant is
the number and position of vortex lines within the condensate
since the precession depends linearly on $\langle L_z \rangle $.
In a condensate of finite size, each vortex line is only
associated with $\hbar$ of angular momentum per particle if it is
exactly centred. The angular momentum associated with an off
centre vortex in an axially-symmetric, harmonically trapped
Thomas-Fermi condensate is \cite{Montserrat,Fett}
\begin{equation}
\langle L_z \rangle = N \hbar \left (
1-\frac{d^2}{R{^2}{_{\perp}}} \right ) ^{5/2} \label{vortexposn}
\end{equation}
where $d$ is the radial position of the vortex and $R_{\perp}$ is
the radial condensate size. Using our second imaging system, that
looks along the axis of rotation (z axis), we were able to check
that $\sim 90 \%$ of runs started with a single, clearly visible
vortex positioned within a third of the condensate radius from the
centre.

Immediately after making a vortex, the TOP trap was suddenly
tilted to excite either the XZ or YZ scissors mode. A detailed
description of the tilting procedure is given in \cite{Marago} but
in summary we apply an additional magnetic field to the TOP trap
in the z direction, oscillating in phase with one of the radial
TOP bias-field components, $B_x$ or $B_y$, to excite either the XZ
or YZ scissors mode respectively. The amplitude of this field was
0.55G, which combined with a 2G radial bias field tilts the trap
by 4.4 degrees and hence excites a scissors oscillation of the
same amplitude about the new tilted equilibrium position.

 After allowing the oscillation to evolve for a variable time in
the trap, we release it and destructively image along the y
direction after $12 \,$ms of expansion. By fitting a tilted
parabolic density distribution to the image, we can extract the
angle of the cloud and thus gradually build up a plot of the
scissors oscillation as a function of evolution time. The
visibility of the fast scissors oscillation depends on the angle
of the cloud projected on the XZ plane (the plane perpendicular to
the imaging direction) and hence varies at the slow precession
frequency $\Omega$. If the oscillation is in the XZ plane then the
projected amplitude is maximum and if it is in the YZ plane then
the projected amplitude is zero. By plotting the scissors
oscillation as a function of evolution time, we observe the slowly
oscillating visibility and hence extract the precession rate.

\begin{figure}
\begin{center}\mbox{ \epsfxsize 2in\epsfbox{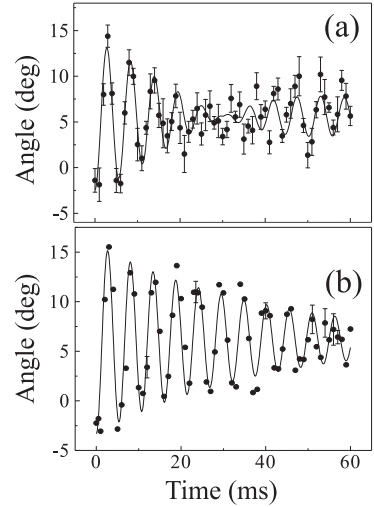}}\end{center}
\caption{The angle of the cloud projected on the XZ plane when the
scissors mode is initially excited in the XZ plane, in (a) with a
vortex and in (b) without a vortex. In (a) each data point is the
mean of 5 runs, with the standard error on each point shown. The
solid line is the fitted function given in eqn.~\ref{fitXZ}. In
(b) most data points are an average of 2 runs, occasionally 5 runs
were taken and the standard error is shown for these points for
comparison with (a). }\label{XZ}
\end{figure}

Figure~\ref{XZ} shows this plot when the scissors mode was
originally excited in the XZ plane, perpendicular to the imaging
direction. In (a) the condensate contained a vortex whilst (b) is
a control run using condensates that did not contain a vortex. The
fitting function used for each was

\begin{equation}
\theta  = \theta_{eq} + \theta_0 \, |\mathrm{cos} \Omega t| \,
(\mathrm{cos} \, \omega_{sc} t) \, e^{- \gamma t}
 \label{fitXZ}
\end{equation}
with $\Omega$ set to zero for fig.~\ref{XZ}b. In both cases the
fast scissors oscillation is clearly visible and the fitted values
of $\omega_{sc} / 2 \pi $ of (a) $179 \,$Hz and (b) $186 \,$Hz
agree reasonably well with the theoretical value of $177 \,$Hz. In
the presence of a vortex the visibility shrinks rapidly to zero
over 30 ms as the oscillation precesses through $ 90^{\circ}$ to a
plane containing the imaging direction. The oscillation visibility
grows again after a further $ 90^{\circ}$ precession. In the limit
of small tilt angles the variation in oscillation visibility is
represented by the $ | \mathrm{cos} \Omega t|$ term in
eqn.~\ref{fitXZ}. Note that $2 \pi / \Omega $ is the time for a
full $2 \pi$ rotation and hence we expect the visibility to fall
from maximum to zero in a quarter period, $\pi /2 \Omega$. The
fitted value of $\Omega / 2\pi = 8.3 \pm 0.7 \,$Hz. From
eqn.~\ref{precession} this gives an angular momentum per particle,
$ \langle l_z \rangle $, of 1.14 $\hbar$ $\pm$ 0.19$\hbar$ for N =
19,000 $\pm$ 4000 atoms. In fig.~\ref{XZ}a, each data point was
taken 5 times and the mean and standard deviation is plotted. This
averaging was necessary because the slight shot-to-shot variation
in the starting conditions, produces slightly different precession
rates.

The revived amplitude is smaller than the initial amplitude due to
Landau damping, which occurs at a rate of $\gamma = 23 \pm 7 \,$Hz
from the exponential decay term in eqn.~\ref{fitXZ}. Damping also
occurs at a similar rate of $\gamma = 25 \pm 5 \,$Hz in the
control run, fig.~\ref{XZ}b, without the presence of a vortex.
Note that in (b) the condensate underwent the same spinning up
procedure but at a trap rotation rate of $35 \,$Hz, just too slow
to create vortices. This ensured that in both cases the
condensates were at the same temperature and hence had comparable
Landau damping rates. The damping rates of approximately $24 \,$Hz
at a temperature of $0.5 \, T_c$ agree well with the data about
the temperature dependence of the scissors mode published in
\cite{Marago2}. The control plot also confirmed that an axially
symmetric condensate must have $L_z = 0$ unless a vortex line is
present and hence the vortex is essential for precession.

\begin{figure}
\begin{center}\mbox{ \epsfxsize 2in\epsfbox{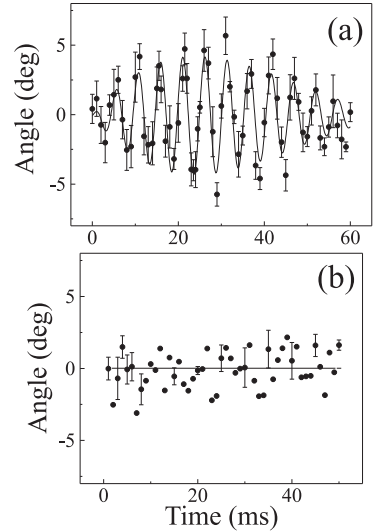}}\end{center}
\caption{As for fig.~\ref{XZ} but with the scissors mode initially
excited in the YZ plane, and eqn.~\ref{fitYZ} as the fitting
function (solid line).} \label{YZ}
\end{figure}

Figure~\ref{YZ} shows the same experiment but with the scissors
mode intially excited along the imaging direction so that the
initial visibility is zero. The appropriate fitting function in
this case is

\begin{equation}
\theta  = \theta_{eq} + \theta_0 \, |\mathrm{sin} \Omega t| \,
(\mathrm{cos} \omega_{sc} t) \, e^{- \gamma t}
 \label{fitYZ}
\end{equation}
There was insufficient data to fit the Landau damping rate
accurately in this case and so the value of $\gamma$ was fixed at
$24.2 \,$Hz, as determined from the data of fig.~\ref{XZ}. In the
presence of a vortex (fig.~\ref{YZ}a) the visibility of the
scissors oscillation grows as the oscillation plane rotates
through $90^{\circ}$ to the XZ plane. This \textbf{growth} of an
oscillation is perhaps a more significant proof of precession than
the initial decrease of amplitude in fig.~\ref{XZ}a, since it
cannot be explained by any damping effect. The precession rate
from fig.~\ref{YZ}a is $7.2 \pm 0.6 \,$Hz, which agrees within the
stated errors with the precession rate of fig.~\ref{XZ}a and gives
$ \langle l_z \rangle  = 0.99 \hbar \pm 0.17 \hbar$. In
fig.~\ref{YZ}b there was no vortex and so the oscillation remained
in the YZ plane, with zero angle projected onto the XZ direction.

Note that the mean angles in fig.~\ref{XZ}a and fig.~\ref{YZ}a are
different. This mean angle corresponds to the trap angle (the
cloud angle in equilibrium) in the visible XZ plane. In
fig.~\ref{XZ} the trap tilt occurs in the XZ plane and so this
mean angle is $\theta_{eq}$, whereas in fig.~\ref{YZ} the tilt is
in the YZ plane, and so the mean angle in the imaging plane is
zero.

Combining the results for the XZ and YZ gyroscope experiments, we
measure the angular momentum per particle associated with a vortex
line to be $1.07 \hbar \pm 0.18 \hbar$. This is in excellent
agreement with the value of $\hbar$ per particle predicted by
quantum mechanics. It is interesting to note that we deduce an
angular momentum per particle slightly greater than $\hbar$, even
though only one vortex is visible. Given that this vortex is not
always perfectly centred, one might expect to observe a value for
$\Omega$ and hence $ \langle L_z \rangle $ that is slightly lower
than the theoretical value.

One explanation is that under conditions which {\it reliably}
produce a single, centred vortex, vortices are also created at the
edge of the cloud, which make a small additional contribution to
the angular momentum (eqn.~\ref{vortexposn}) but are not
observable. This explanation is in agreement with the results of
Chevy {\it et al.} in \cite{Chevy}.

\begin{figure}
\begin{center}\mbox{ \epsfxsize 2in\epsfbox{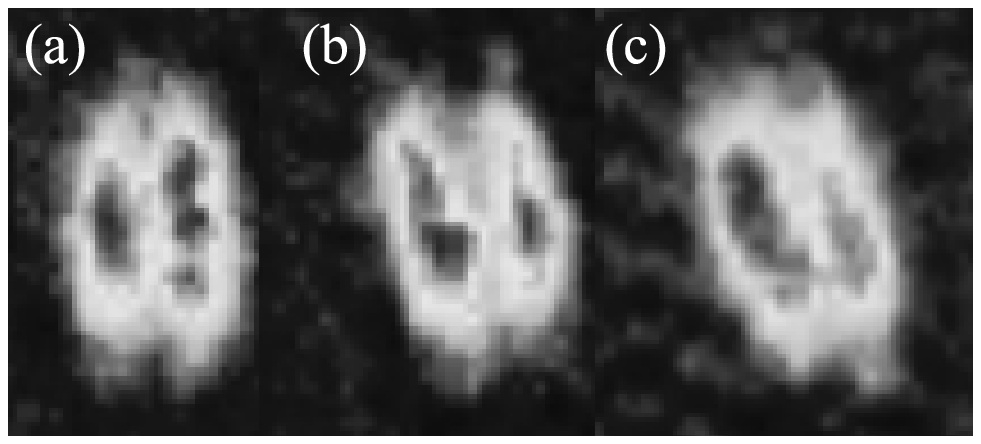}}\end{center}
\caption{Absorption images taken along the y direction after $12
\,$ms of expansion show the vortex core following the angle of the
axis of the condensate during the gyroscope motion. These images
were taken with a smaller condensate density ($N \leq 10,000$), so
that the vortex core (of radius $(8 \pi n a)^{-1/2}$) was large
enough to have a significant effect on the integrated absorption
profile. }\label{sidepic}
\end{figure}

Finally we discuss the motion of the vortex core, prompted by the
suggestion in \cite{Stringari} that it might exactly follow the
axis of the condensate. Images taken perpendicular to the vortex
core were necessary to show the angle of the core relative to the
axis of the condensate. However in a dense condensate the vortex
core is too small to have a discernable effect on the integrated
absorption profile. Thus we reduced the number of atoms in the
condensate to $\leq 10,000$ to obtain the images in
fig.~\ref{sidepic}. Vortices were clearly visible on $50 \%$ of
the experimental runs and within the limits of the imaging system
resolution and pixel to pixel noise, it appears that the vortex
line is aligned with the condensate axis in $95\%$ of these shots.

In conclusion, we have observed a superfluid gyroscope effect in a
trapped BEC, in which a single quantum of circulation (a single
vortex line) affects the bulk scissors motion of the condensate
and causes it to precess. By observing the precession rate we are
able to measure the angular momentum associated with the vortex
line. Our result of $1.07 (\pm 0.18) N \hbar$ agrees well with $N
\hbar$, the quantum of angular momentum that is predicted to be
associated with each centred vortex line.

We acknowledge support from the EPSRC, St. John's College, Oxford
(G.H.), Christ Church College, Oxford (E.H.). We also thank Graham
Quelch for his technical support.



\end{document}